\documentclass[prl,twocolumn,showpacs]{revtex4}

\usepackage{amsmath}
\usepackage{mathrsfs}
\usepackage{graphicx}

\newcommand{\bracket}[--2]{\langle\hspace{0.5pt}#1\hspace{0.5pt}|\hspace{0.5pt}
    #2\hspace{0.5pt}\rangle}
\renewcommand{\v}[1]{\ensuremath{\mathbf{#1}}}
\begin{document}

\title{Gaussian potentials facilitate access to quantum Hall states in 
rotating Bose gases}

\author{Alexis G. Morris}
\author{David L. Feder}
\affiliation{Department of Physics and Astronomy and Institute for Quantum
Information Science, University of Calgary, Calgary, Alberta, Canada T2N 1N4}

\begin{abstract}
Through exact numerical diagonalization for small numbers of atoms, we show
that it is possible to access quantum Hall states in harmonically confined
Bose gases at rotation frequencies well below the centrifugal limit by
applying a repulsive Gaussian potential at the trap center. The main idea is
to reduce or eliminate the effective trapping frequency in regions where the
particle density is appreciable. The critical rotation frequency required to
obtain the bosonic Laughlin state can be fixed at an experimentally accessible
value by choosing an applied Gaussian whose amplitude increases linearly with
the number of atoms while its width increases as the square root.
\end{abstract}


\maketitle

For almost 30 years since its initial observation in a two-dimensional electron
gas~\cite{Klitzing:1980}, the quantum Hall (QH) effect has continued to be
the subject of intense research. Much recent theoretical work indicates that
similar incompressible states should also exist in ultracold trapped atomic
gases~\cite{Regnault:2004,Regnault:2006}; the most intriguing possibility
would be neutral bosons with tunable two-particle interactions. Many of the
proposals exploit the formal equivalence of the Hamiltonian for a
two-dimensional interacting electron gas in the presence of an external
transverse magnetic field with that describing a neutral interacting gas
subject to rotation~\cite{Cooper:1999,Regnault:2003}. Other theoretical approaches make use of internal
degrees of freedom to create artificial magnetic
fields~\cite{Jaksch:2003,Mueller:2004,Sorensen:2005,Ruseckas:2005}.

In addition to their inherent interest, fractional QH states may also be used
to perform intrinsically fault-tolerant quantum computing. The topological
properties of quasiparticle excitations with non-Abelian statistics could be
used to generate a universal set of multi-qubit gates that are protected
against the deleterious effects of decoherence~\cite{Kitaev:1997} by braiding
them around each other in specific patterns~\cite{Bonesteel:2005}. The
QH state with filling factor $\nu=5/2$ in electronic systems is widely
believed to be described by a Moore-Read wavefunction which possesses the
required properties, though recent work has cast some doubt on
this~\cite{Toke:2006,Toke:2007}. The equivalent state in rotating Bose gases
(RBGs) is expected at $\nu=1$~\cite{Regnault:2003}. A distinct advantage of ultracold atomic gases
for topological quantum computing is that quasiparticle creation and
manipulation could be as straightforward as shining blue-detuned lasers into
the gas and moving them around each other~\cite{Paredes:2002}. 

Unfortunately, the QH effect has not yet been observed in RBGs. Incompressible
states result when the number of quantized vortices $N_v$ (the analog of
quantized flux in electronic systems) is comparable to the number of particles
$N$, so that the filling factor $\nu\equiv N/N_v\sim 1$. This
strongly correlated regime generally requires either very low particle number, or very high rotation frequencies
approaching the centrifugal limit of the confining harmonic 
trap~\cite{Rosenbusch:2002}. The first issue can be addressed by adding a deep
one-dimensional optical lattice oriented along the axis of rotation, making an
array of disconnected pancake-shaped wells~\cite{Hadzibabic:2004}, each containing a small number of atoms (conceivably on the order of $10^3$),
of which all but one or two can be eliminated~\cite{Stock:2005:2}. The second
issue can be partly resolved by applying an approximately quartic potential to
the harmonic trap (in practice by adding a repulsive Gaussian
potential)~\cite{Stock:2005}, though at high rotation frequencies this
appears to favor weakly correlated `giant vortex' states~\cite{Aftalion:2004}.

Using exact calculations for small numbers of particles, we demonstrate that
the application of a repulsive Gaussian potential with the appropriate
parameters can yield bosonic fractional quantum Hall states in RBGs at
experimentally accessible rotation frequencies. A simple model agrees
closely with the numerical results. To favor the Laughlin state, the combined 
potentials (harmonic trap, centrifugal term, and external Gaussian) should be 
flat over the spatial region where the particle density is appreciable. This 
requires a Gaussian whose amplitude increases linearly with the number of 
bosons $N$ (with a small coefficient) while the width scales as $\sqrt{N}$.

The Bose gas at zero temperature is harmonically trapped in an axisymmetric
potential with radial and axial trapping frequencies $\omega$ and $\omega_z$,
respectively, and is rotated around the $z$-axis at a rate $\tilde\Omega$.
Imposing tight confinement along the $z$-axis ($\omega_z\gg\omega$), all
motion along this direction is frozen out and the cloud is
quasi-two-dimensional, with each particle's position denoted 
${\bf r}_i=(\rho_i,\phi_i)$. An external Gaussian potential of the form
$V_\text{ext}=\tilde\gamma e^{-\rho^2/2\tilde\sigma^2}$ is included to
represent a repulsive blue-detuned laser that is being shined upon the center
of the rotating Bose gas, with tunable amplitude $\tilde\gamma$ and width
$\tilde\sigma$. The bosons interact via the usual delta-function
pseudopotential with a two-dimensional coupling constant
$\tilde{g}=\sqrt{8\pi}\hbar\omega \ell^2 a/\ell_z$, where variables
$\ell=\sqrt{\hbar/M\omega}$ and $\ell_z=\sqrt{\hbar/M\omega_z}$ are the
characteristic oscillator lengths along the radial and axial directions,
respectively, and $a$ is the three-dimensional scattering length.
In the rotating frame the Hamiltonian is thus
\begin{eqnarray}
H&=&\sum_{i=1}^N \bigg[\frac{1}{2M}\left(-i\hbar\v{\nabla}_i
-M\tilde{\v{\Omega}}\times{\bf r}_i\right)^2 + \frac{M}{2}\omega_z^2 z_i^2
\label{H} \\
&+&\frac{M}{2}\left(\omega^2-\tilde{\Omega}^2\right)\rho_i^2
+\tilde\gamma e^{-\rho_i^2/2\tilde\sigma^2}\bigg] 
+ \tilde g\sum_{i<j} \delta({\bf r}_i-{\bf r}_j),\nonumber
\end{eqnarray}
where $N$ is the number of bosons and $M$ their mass. The energy spectrum of
the RBG can then be calculated by exact
diagonalization of the Hamiltonian $H$, for various values of the parameters
$\Omega\equiv\tilde\Omega/\omega$,  $g\equiv\tilde g/\hbar\omega\ell^2$,
$\gamma\equiv\tilde\gamma/\hbar\omega$ and $\sigma\equiv\tilde\sigma/\ell$.
For more details about the diagonalization process, see Ref.~\cite{Morris:2006}.

\begin{figure}[t]
\begin{center}
\includegraphics[width=0.48\textwidth]{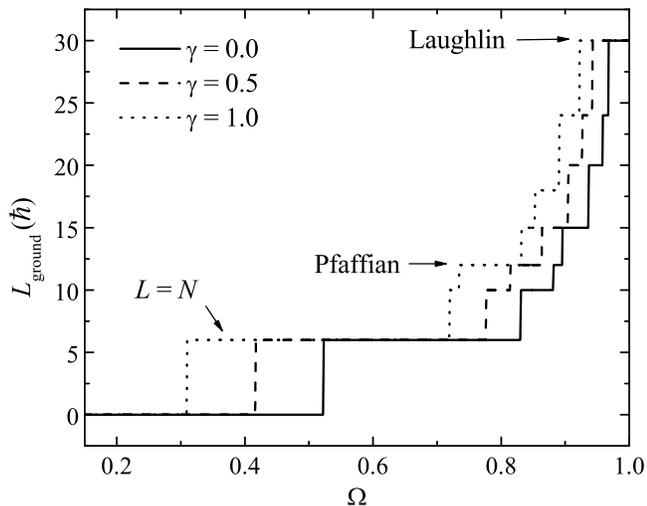} 
\caption{The yrast lines for the six-particle RBG for different external
potential intensities $\gamma$. Parameter values are $\sigma=2.0$
and $g=0.1$.}
\label{yrast}
\end{center}
\end{figure}

A Bose condensate is irrotational and thus can only acquire angular momentum
through the nucleation of vortices~\cite{Madison:2000}. For small rotation
rates, the condensate's ground state has zero angular momentum. As $\Omega$
increases and more vortices penetrate the cloud, the system undergoes a series
of transitions from states with low to high angular momentum~\cite{Wilkin:2000},
eventually yielding incompressible QH-like states as $\Omega\to 1$.
Fig.~\ref{yrast} shows the corresponding `yrast line'~\cite{Bertsch:1999} for
six particles.  Currently, rotation rates in excess of $\Omega=0.99$ have been
achieved in the laboratory~\cite{Schweikhard:2004}, though with current
experimental conditions the filling factor $\nu$ is still an order of
magnitude too high to reach the QH limit.
 
\setlength{\abovecaptionskip}{3pt}
 
The addition of $V_\text{ext}$ has a profound effect on the RBG's energy
spectrum since a narrow gaussian primarily increases the energy of levels with
small angular momentum $L$, because particles in these states are situated
closer to the origin. Consequently, including a repulsive Gaussian favors
ground states with higher $L$, lowering the various critical rotation
frequencies. As shown in Fig.~\ref{yrast}, increasing $\gamma$ at constant
$\sigma$, $g$, and $\Omega$, changes the ground state from a Bose condensate
with $L=0$ to the `single vortex'
state with $L=N$,
eventually to the so-called Pfaffian state at $L=N(N-2)/2$,
and ultimately to the Laughlin state at $L=N(N-1)$~\cite{Wilkin:2000}.

While higher angular momentum states may be favored by the presence of 
$V_\text{ext}$, and the resulting particle density might look similar as in
its absence, it is not obvious that the strongly correlated nature of these
states is preserved. To address this issue one needs to consider
higher-order correlation functions. In its normalized form, the pair (also
known as the density-density or second-order) correlation function is defined
as:
\begin{equation}
g_2({\bf r},{\bf r}')=\frac{\left<\hat\psi^\dagger(\v{r})\hat\psi^\dagger
(\v{r'})\hat\psi(\v{r'})\hat\psi(\v{r})\right>}{\left<\hat\psi^\dagger(\v{r'})
\hat\psi(\v{r'})\right>\left<\hat\psi^\dagger(\v{r})\hat\psi(\v{r})\right>,}
\label{g2}
\end{equation}  
where $\hat\psi^\dagger(\v{r})$ and $\hat\psi(\v{r})$ are the bosonic creation
and destruction field operators, respectively, and the expectation value 
$\langle\cdots\rangle$ is taken with respect to the ground state. The
density-density correlation function can be interpreted as the probability of
a particle being at position ${\bf r}$ when another is at position ${\bf r'}$,
and has been used to probe for pairing in a superfluid Fermi
gas~\cite{Greiner:2004,Lobo:2006} and for density-wave order in bosons
confined within deep optical
lattices~\cite{Folling:2005,Duan:2006,Spielman:2007}.

\begin{figure}[t]
\begin{center}
\includegraphics[width=0.48\textwidth]{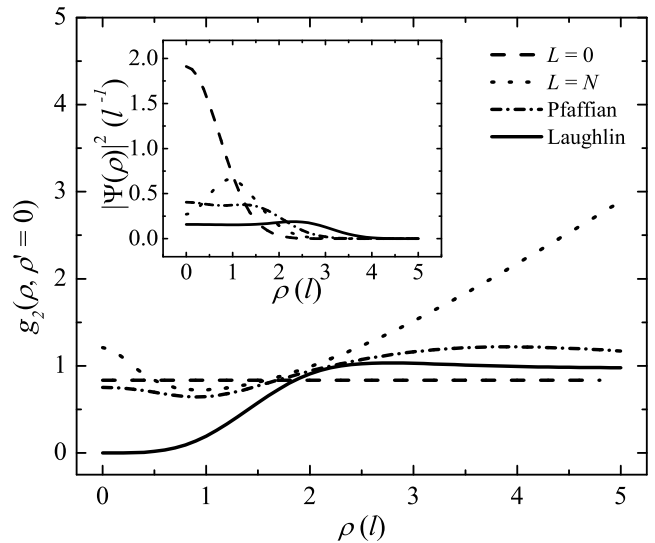} 
\caption{Second order radial pair correlation function for the $L=0$, $L=N$,
Pfaffian, and Laughlin states of a six-particle RBG, with one particle fixed
at the origin ($\rho'=0$), in the absence of any external potential ($\gamma=0$). The inset shows the corresponding densities.}
\label{pc1}
\end{center}
\end{figure}

As shown in Fig.~\ref{pc1}, the Laughlin state has a particularly distinctive
radial pair correlation function $g_2(\rho,\rho')$ (the $\phi$ dependence is
not accessible due to the cylindrical symmetry of our system), which would
make its experimental detection
unambiguous. The Laughlin wavefunction is an exact solution of the many-body
Hamiltonian~(\ref{H}) when $\Omega=1$, and is defined as
\begin{equation}
\Psi_{\rm Laughlin}={\mathcal N}\prod_{i<j}^N (\v{z}_i-\v{z}_j)^2
e^{-\sum_{i}^N|z_j|^2/4\ell_c^2},
\label{Laughlin}
\end{equation}
where ${\bf z}_j=\rho_je^{i\phi_j}$ is the complex coordinate of particle $j$,
and ${\mathcal N}$ is a normalization constant. The characteristic length
$\ell_c=\sqrt{\hbar/m\omega_c}$ is defined in terms of the cyclotron frequency
$\omega_c=2\tilde{\Omega}$, so that $\ell_c\approx\ell/\sqrt{2}$ when
$\tilde{\Omega}\sim\omega$. Because of the Jastrow-type prefactor, the value of 
$g_2({\bf r},{\bf r}')$ is expected to vanish when two particles overlap
({\bf r}={\bf r}'). In this respect the Laughlin state is most different from
the weakly correlated Bose-Einstein condensates at low $L$, for which
$g_2(\v{r},\v{r}')$ is a constant approaching unity as $N\to\infty$ for any
$\v{r}$ and $\v{r}'$.

\begin{figure}[t]
\begin{center}
\includegraphics[width=0.45\textwidth]{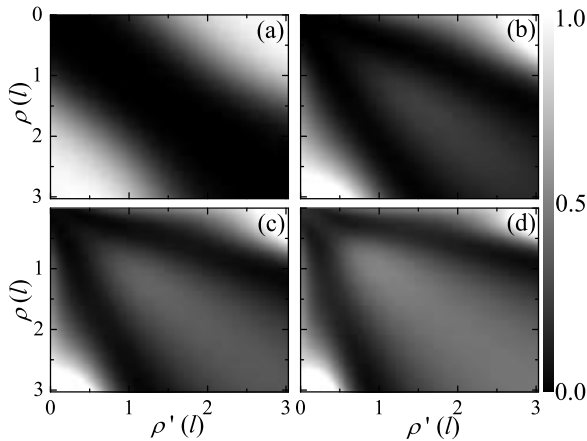} 
\caption{Radial pair correlation for the five-particle Laughlin state, in the
absence of any external potential (a).  As the external potential amplitude 
$\gamma$ is increased to 
$\gamma=0.25$ (b), $\gamma=0.5$ (c), and $\gamma=1.0$ (d), the likelyhood of overlapping particles increases suggesting that the ground state is no longer the Laughlin. Parameters correspond to $g=0.1$, and $\sigma=1.4$.}
\label{pc2}
\end{center}
\end{figure}

Fig.~\ref{pc2} shows $g_2(\rho,\rho')$ with $\rho,\rho'\le 3\ell$,
for the ground state at $\Omega=1$ under the influence of different external
potential intensities $\gamma$ and fixed $\sigma=1.4$. If $V_{\rm ext}$
had no influence on the Laughlin state, the whole diagonal would remain zero.
Since this is evidently not the case, it is critical to choose the parameters
$\gamma$ and $\sigma$ judiciously to best preserve the fundamental
characteristics of the Laughlin state, particularly its pair correlations.

\begin{figure}[!t]
\begin{center}
\includegraphics[width=0.45\textwidth]{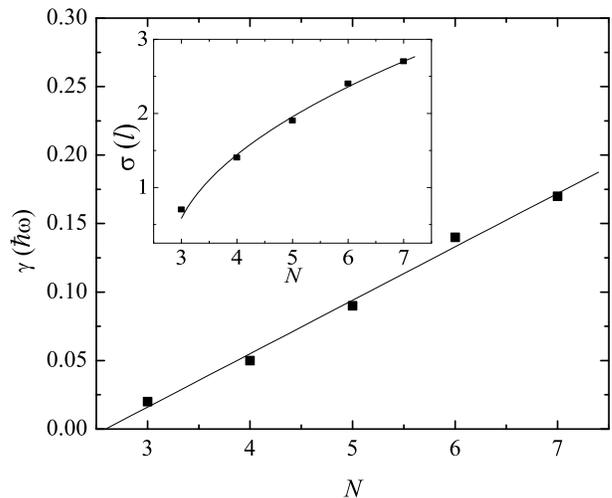} 
\caption{Optimized magnitude ($\gamma$) and width ($\sigma$) of the external
gaussian potential such that the RBG ground state is the Laughlin state for
$\Omega_c= 0.99$ and $g=0.1$. The lines are best fits, corresponding to
$\gamma=(0.039\pm 0.002)N-(0.10\pm 0.01)$ and
$\sigma=\sqrt{(1.74\pm 0.06)N-(4.9\pm 0.3)}$.}
\label{rot99}
\end{center}
\end{figure}

For each $N$ the parameters defining $V_{\rm ext}$ were varied in the range
$\{0,20\}$ in their appropriate units, with $\sigma$ and $\gamma$ in
increments of $0.1$ and $0.01$, respectively. For each case, the location
of the radial pair correlation maximum $g_2(\rho_m,\rho_m)$ along the diagonal was
determined, and any parameters for which the maximum exceeded $0.01$ were
discarded. The values of the parameters $\gamma$ and $\sigma$ were
minimized, subject to the condition that the RBG ground state is the Laughlin
state for some chosen, fixed, value of the rotation frequency $\Omega=\Omega_L<1$.
Results are plotted in Fig.~\ref{rot99} for $\Omega_L=0.99$ and $g=0.1$. These indicate that a linear increase in $\gamma$
with $N$ and $\sigma\propto\sqrt{N}$ are sufficient to attain the Laughlin
state at $\Omega_L=0.99$. Extrapolating from the small-$N$ data in
Fig.~\ref{rot99}, one obtains $\gamma=39 \pm 2$ and $\sigma=42\pm 1$ for
$N=10^3$. Even for $N=10^4$, the values are not completely unreasonable:
$\gamma\approx 400$ and $\sigma\approx 140$.  For a choice of $\omega=100$Hz, this 
corresponds to an amplitude of $\tilde \gamma/h = 40$kHz, while optical lattices with 
depths of 50kHz are readily produced \cite{Hadzibabic:2004}.

The qualitative behavior described above was found to be quite general. The
numerical calculations were repeated using two and three landau Levels, and 
the results were identical to those obtained within the lowest Landau level
approximation for interaction strengths $0\leq g\leq 1$. The optimal values of
the Gaussian parameters depend on both $g$ and $\Omega$, but their functional
dependence on $N$ is unchanged. When $g$ is increased, it is easier to reach
the quantum Hall states \cite{Morris:2006}, and therefore the values of
$\sigma$ and $\gamma$ are smaller.  Similarly, should a lower transition
rotation frequency $\Omega_L$ be desired keeping $g$ constant, then the
Gaussian needs to be higher and wider. For example, for $\Omega_L=0.95$
the best fits to the data yield $\gamma=(1.05\pm 0.05)N-(2.3\pm 0.3)$ and
$\sigma=\sqrt{(10.2\pm 0.5)N-(23\pm 3)}$ which yields extrapolated values
$\gamma=1050\pm 50$ and $\sigma=101\pm 3 $ for $N=1000$. Evidently, a very low
transition frequency for the Laughlin state is not feasible with this
technique.

To further validate the pair correlation function as a figure of merit, one
requires high overlap $\bracket{\Psi_\text{Laugh}}{\Psi}$ between the exact
Laughlin state and the one obtained in the presence of the external potential.
For all $N$ considered, the overlap integral exceeded $0.99$. Plotting the
overlap as a function of $1/N$ shows an asymptotic behavior, with a limiting
value for large $N$ of 98.5\% for the $\Omega_L=0.99$ data shown in 
Fig.~\ref{rot99}. Thus, for the optimized values of $\gamma$ and $\sigma$, the
state remains the bosonic fractional quantum Hall Laughlin state.

A simple model accounts for the success of applied Gaussians in reducing the 
critical frequency for the Laughlin state: the combination of the external
potentials and the centrifugal barrier should mimic a constant potential in
the spatial region where the particle density is appreciable. For large
$\sigma$, the Gaussian can be expanded in a Taylor series and the radial
contribution to the single-particle potential $V(\rho_i)$ in Eq.~(\ref{H}) can
be approximated as
\begin{equation}
\frac{V(\rho_i)}{\hbar\omega}\approx\gamma+\frac{1}{2}\left[1-\Omega^2
-\frac{\gamma}{\sigma^2}\right]\frac{\rho_i^2}{\ell^2}
+\frac{\rho_i^4}{8(\sigma\ell)^4}+{\mathcal O}\left(\rho_i^6\right).
\label{Veff}
\end{equation}
Enforcing that the term in the square brackets vanishes when $\Omega=\Omega_L$, one obtains the
relationship between the parameters of the Gaussian:
$\gamma=\left(1-\Omega_L^2\right)\sigma^2$. To ensure that the Laughlin state
experiences little of the quartic contribution in Eq.~(\ref{Veff}), one
requires that the RMS width of the Laughlin state satisfy 
$\rho_{\rm RMS}=\sqrt{\langle\rho^2\rangle}\lesssim\sigma$ for all $N$. The
spread of $\Psi_{\rm Laughlin}$ can be estimated by noting that the Jastrow
prefactor in the definition~(\ref{Laughlin}) can be written as a polynomial
of order $N(N-1)$ in the variables $\rho_i$ with complex coefficients. Thus
we obtain $\Psi_{\rm Laughlin}\approx{\mathcal N}e^{iN(N-1)\phi}\rho^{N(N-1)}
e^{-N\rho^2}$. Straightforward integration yields 
$\rho_{\rm RMS}=\sqrt{N-1+(1/N)}\approx\sqrt{N}$ for $N\gg 1$. Indeed,
direct evaluation using (\ref{Laughlin}) for small $N$ gives exactly
$\rho_{\rm RMS}=\sqrt{N}$. Enforcing the condition that
$\rho_{\rm rms}\sim\sigma$, one obtains $\sigma\sim\sqrt{N}$ and
$\gamma\sim N$, in agreement with the results shown in Fig.~\ref{rot99}.

The best fits to the numerical data are consistent with these scaling
relations. 
For $g=0.1$ and $\Omega_L=0.99$, one obtains
$\sigma=\sqrt{1.74N-4.9}$, yielding an optimal value for the
amplitude of $\gamma=(1.74N-4.9)(1-\Omega_L^2)$, from which the correct slope and y-intercept values are found. Intriguingly, for $\Omega_L=0.95$, one
rather obtains $\sigma=\sqrt{10.2N-23}$, so that the
Gaussian width must be considerably wider than the mean spread of the Laughlin
density as the desired rotation frequency is lowered.
It is important to note that these choices for the Gaussian
parameters facilitate access to not only the Laughlin state, but to all ground
states: the effective trap frequency is reduced from $\sqrt{1-\Omega^2}$ to
$\sqrt{\Omega_L^2-\Omega^2}$, so all transition frequencies are accordingly
reduced.

In conclusion, we have demonstrated that it is possible to attain the 
Bose-Laughlin quantum Hall state, and thus any other quantum Hall state, by
applying an external Gaussian potential to a harmonically trapped rotating
Bose gas. With experimentally reasonable choices of Gaussian amplitude and
width, number of particles, and effective interaction strength, the transition
to a Laughlin state can occur for the experimentally accessible rotation rate
of 0.99$\omega$ or lower. These results suggest that quantum Hall physics in
neutral, rotating Bose systems are currently within experimental reach.

\acknowledgments

The authors are grateful to M.~Oberthaler and H.~Stoof for their insightful
comments. This work was supported by the Natural Sciences and Engineering
Research Council of Canada, the Canada Foundation for Innovation
and Alberta's Informatics Circle of Research and Excellence.

\bibliographystyle{aip}
\bibliography{externalpot}

\end{document}